\documentclass[twoside,fleqn]{article}
\usepackage{graphicx}
\usepackage{espcrc2}

\usepackage{graphics}

\newcommand{\AmS}{{\protect\the\textfont2
 
  A\kern-.1667em\lower.5ex\hbox{M}\kern-.125emS}}

\title{\mbox{$\pi\pi$} scattering amplitudes constrained by Roy's equations\thanks{Talk presented at QCD05 conference, Montpellier, July 2005}\thanks{This work has been performed in the framework of the IN2P3-Polish Laboratories Convention (project number 99-97)}}

\author{R. Kami\'nski$^a$, L. Le\'sniak$^a$ and B. Loiseau$^b$\\
\addressmark{$^a$ Department of Theoretical Physics, The Henryk 
Niewodnicza\'nski Institute of Nuclear Physics,
Polish Academy of Sciences, 31-342 Krak\'ow, Poland\\      
$^b$ Laboratoire de Physique Nucl\'eaire et de Hautes 
\'Energies\thanks{Unit\'e de Recherche des Universit\'es Paris 6 et Paris 7, associ\'ee au CNRS}, Groupe Th\'eorie, Univ. P. \& M. Curie, 4 Pl. Jussieu, F-75252 Paris, France}}

\begin{document}

\begin{abstract}
The scalar-isoscalar, scalar-isotensor and vector-isovector $\pi\pi$ amplitudes
have been fitted simultaneously to experimental data and to to Roy's equations.
Resulting $\pi\pi$ phase shifts up to 1600 MeV and  near threshold observables have been analyzed.
Only the amplitudes fitted to the "down-flat" set of phase shifts in scalar-isoscalar wave fulfill 
crossing symmetry conditions and can be regarded as physical. 
\end{abstract}

\maketitle

\section{Introduction}

After the new analysis of the $\pi^-\,p_{\uparrow} \to \pi^+\,\pi^-\,n$ reaction on polarized target in 1997, the long standing two-fold ``up-down'' 
ambiguity in scalar-isoscalar $\pi\pi$ amplitudes below 1000 MeV reappeared as a four-fold one \cite{kam97}. 
Two of these solutions, called ``steep'',  violate unitarity and can be rejected as non-physical \cite{kam97,kam99}. 
The phase shifts for the two left ``flat'' data sets are presented in Fig.~1. 
\begin{figure}[htb]
\includegraphics*[width=17.7pc]{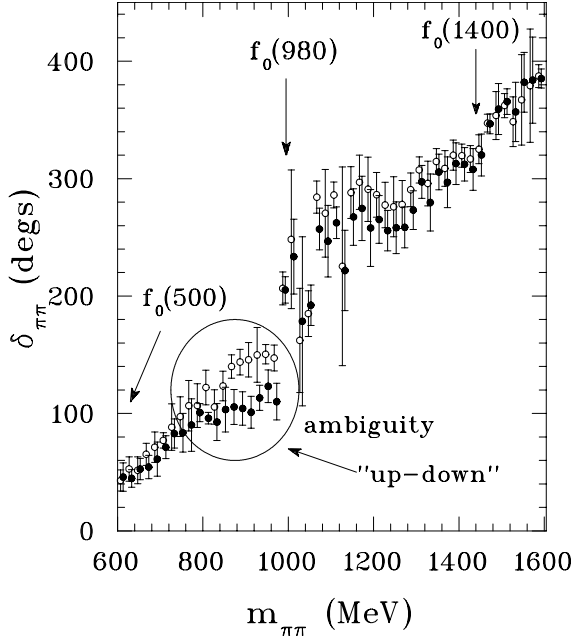}

\vspace{-.5cm}

\caption{
Scalar-isoscalar phase shifts \cite{kam97} for the solutions ``up-flat'' (open circles) and ``down-flat'' (full circles);
the round curve indicates the region where the ``up-down'' ambiguity leads to the largest differences between 
the two solutions.}
\label{fig:ambiguity}
\end{figure}
The biggest differences are in the effective two-pion mass range $m_{\pi\pi}$ between 800 and 970 MeV i.e. near 
the two scalar resonances $f_0(500)$ and $f_0(980)$.
The knowledge of the phase shifts in that region is important for example in the determination of the parameters of these scalar resonances and it is 
crucial for works on dispersion relations for the $\pi\pi$ amplitudes below 1000 MeV \cite{PY}.

\section{Analysis of the data}

To determine whether unitary amplitudes, fitted to the ``flat'' solutions, can be treated as physical ones we have 
checked if they fulfill crossing symmetry conditions \cite{kll03}.
We have used, as input amplitudes in Roy`s equations, the scalar-isoscalar (S0), scalar-isotensor (S2) and vector-isovector (P1) waves 
defined in \cite{kll03,kll97}.
We have performed simultaneous fits to experimental data and to Roy`s equations in the $m_{\pi\pi}$ range from the $\pi\pi$ threshold to 
970 MeV using the $\chi^2$ test for $\chi^2_{tot}=\chi^2_{exp}+\chi^2_{Roy}$ with
\begin{equation}
\chi^2_{exp} =\hspace{-0.2cm} \sum_{I=0,1,2}\hspace{-0.1cm} \sum_{\,\,i=1}^{N_I} 
\left\lbrace \frac{\sin \left[\delta_{\ell}^I \left(s_i\right) 
- \varphi_{\ell}^I \left(s_i\right)\right]}
       {\Delta \varphi_{\ell}^I \left(s_i\right)} \right\rbrace^2,
        \label{eq:chi:exp}
\end{equation}
\begin{equation}
\chi^2_{Roy} \hspace{-0.1cm}= \hspace{-0.3cm}\sum_{I=0,1,2}\hspace{-0.1cm} \sum_{\,\,j=1}^{12} 
\hspace{-0.1cm}	\left\lbrace \frac{Re\,f_{out}^I\left(s_j\right)-Re\,f_{in}^I\left(s_{j}\right)}
	{\Delta f} \right\rbrace^2,
        \label{eq:chi:Roy}
\end{equation}
where $\varphi_{\ell}^I \left(s_i\right)$ and 
$\Delta \varphi_{\ell}^I \left(s_i\right)$ represent the experimental phase 
shifts and their errors, respectively. 
The real parts  $Re\, f_{in}^I(s_i) = \eta\,sin[2\delta^I_{\ell}(s_i)]/(2\,k_i\,s^{1/2}_i)$ 
are expressed as functions of the $\pi\pi$ phase shifts $\delta^I_{\ell}(s_i)$ and
have been calculated under the assumption that the inelasticity $\eta$ is equal to 1 in the studied $m_{\pi\pi}$ range. 
The other real parts, denoted by Re~$f^I_{out}(s_i)$ are the output values calculated from Roy's equations.
We took $\Delta f$ value of $0.5\times 10^{-2}$ to obtain reasonable fits to Roy's equations. 
The $N_I=18$ experimental values of the ``up-flat'' or ``down-flat'' data between 600 and 950~MeV were used in addition to the six data points taken from 
\cite{pislak01}.

The S0-wave amplitudes have been constructed by fitting the three coupled-channel model predictions for phase shifts and inelasticities (see \cite{kll97})
to the data of \cite{kam97}.   
In the S2-wave we have used the data from solution A of Hoogland {\it et. al.} \cite{Hoogland} and in the P1-one the data of Hyams {\it et. al.} \cite{Hyams}.

Below 970~MeV the following Pad\'e representation of the S0 phase shifts has been taken:
\begin{equation}
    \tan\delta_{0}^0(s)= \frac{\sum_{i=0}^{4}\alpha_{2i+1}k^{2i+1}}
     {\Pi_{i=1}^{3}(k^2/\alpha_{2i}-1)},
    \label{eq:Pade}
\end{equation}
where $k=\frac{1}{2}\sqrt{s-4m_{\pi}^2}$ is the  pion  momentum and 
$\alpha_{j}\,(j=1,\ldots,7,9)$ are constant parameters. 
Above 970~MeV and up to 2~GeV our coupled channel model \cite{kll97} amplitude A, fitted to the ``down-flat''
data, and 
the amplitude C, constrained by the ``up-flat'' data, were used.
These two representations, one below 970 MeV and the second above were joined smoothly  at 970 MeV up to their first derivative.
In the fits we have also used the near-threshold phase shifts calculated from the differences $\delta_0^0-\delta_1^1$
obtained in the high statistics $K_{e4}$ decay experiment \cite{pislak01}. 

The parameterization of the S2-wave, using a rank-two separable potential model,  has been described 
in \cite{kll03} where detailed analysis of the present study is presented.

For the $P$-wave, from the $\pi\pi$ threshold up to 970~MeV, we have used an extended Schenk parameterization 
\cite{anan}:
\begin{equation}
\begin{array}{ll}
\tan \,\delta_1^1(s) =
\frac{2}{\sqrt{s}}\,k^3\\
\hspace{0.cm}\times\left(A+Bk^2+Ck^4+Dk^6\right)
\left(\frac{4m_{\pi}^2-s_\rho}{s-s_\rho}\right),
\label{eq:schenk}
\end{array}
\end{equation}
where $A$ is the $P$-wave scattering length and $s_\rho$ is equal to the $\rho$-mass squared. 
Above 970~MeV we took the $K$-matrix parameterization of Hyams et al. \cite{Hyams}. 
The parameters $C$ and $D$ were chosen to join smoothly around 970~MeV the phase shifts given by both parameterizations.

The contributions to Roy's equations from high energies ($m_{\pi\pi} > 2$ GeV) and from higher partial waves 
($l>1$), called driving terms, are composed of contributions from the $f_{2}(1270)$ and $\rho_{3}(1690)$ resonances and 
from the Regge amplitudes for the Pomeron, $\rho$- and $f$-exchanges.
For the $f_{2}(1270)$ and $\rho_{3}(1690)$ we have used the Breit-Wigner parameterization with masses, widths and $\pi\pi$ 
branching ratios taken from \cite{pdg02}.
For the Regge parts we have used formulae of \cite{anan} without the $u$-crossed terms.
In the driving term the  $f_{2}(1270)$ resonance dominates in the scalar-isoscalar wave and the isotensor and isovector waves are mostly 
influenced by the $\rho_{3}(1690)$.
In the isoscalar wave the Regge contributions are more than 10 times smaller than the resonance contributions
but for the isospin 1 and 2 they are of the same order.

\begin{figure}[h!]
\includegraphics*[width=17.7pc]{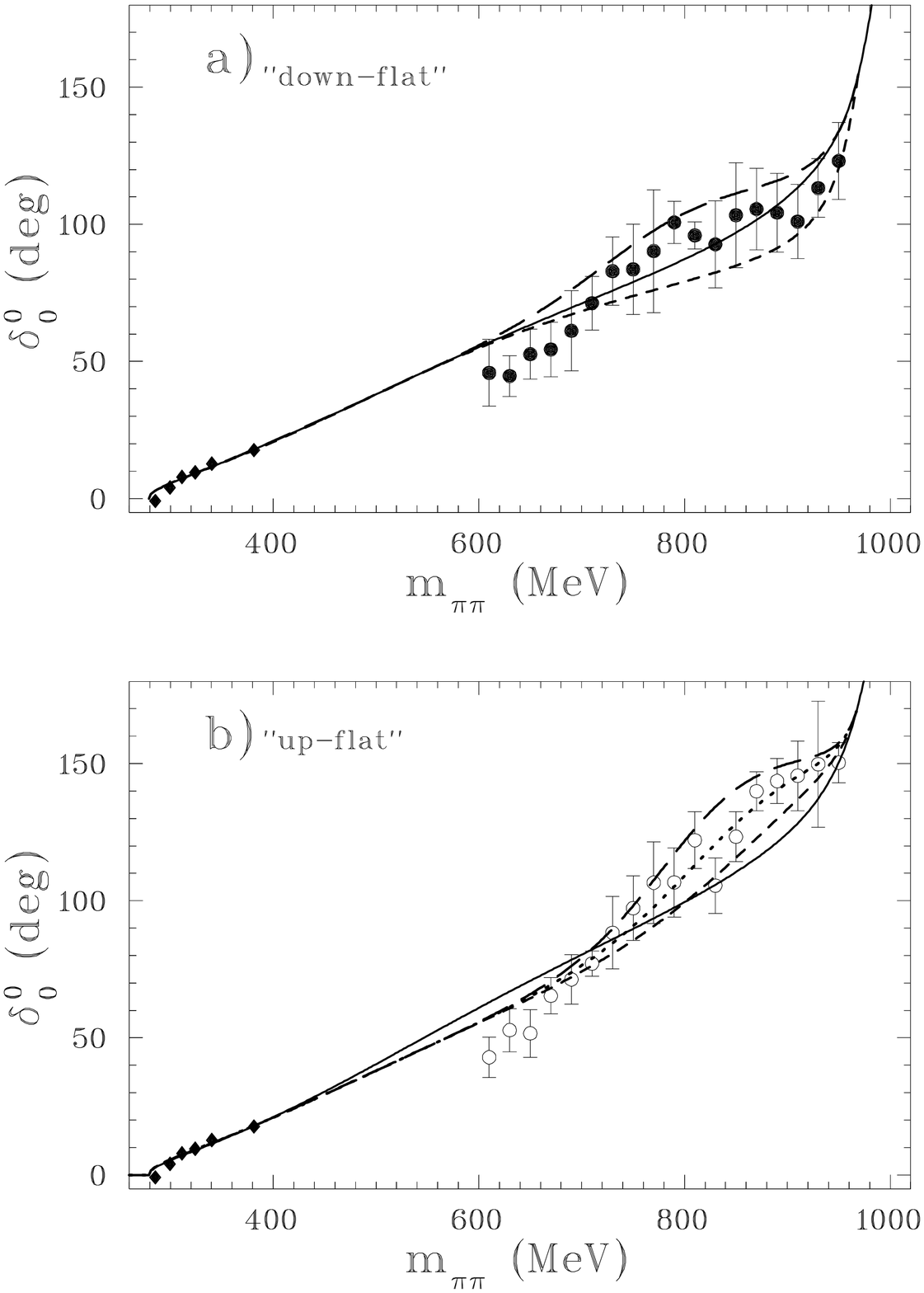}

\vspace{-0.8cm}

\caption{Solid lines: fits to the scalar-isoscalar phase shifts of [1] and to Roy`s equations; dashed lines: fits to 
the data points shifted upwards (called {\it ``upper''} and drawn as long-dashed lines) and downwards 
(called {\it ``lower''} and drawn as short-dashed lines) by their errors. 
Dotted line for the ``up-flat'' solution represents fit to experimental data only.}
\label{fig:bands}
\end{figure}
In Fig.~2a and 2b we present results of fits to the "down-flat" and "up-flat" phase shifts and to Roy's equations (solid
lines). 
In both cases the differences $\mid$~Re~$f^I_{out}$~-~Re~$f^I_{in}$~$\mid$
are of the order of $10^{-3}$ in all three partial waves.
The $\chi^2$ for 18 points between 600 and 970~MeV was 16.6 in the "down-flat" and 46.4 in the "up-flat" cases.
We see in Fig.~2b that the solid line lies distinctly below the "up-flat" data points between 
800 and 970~MeV. 
In contrary, the corresponding line for the "down-flat" case in Fig.~2a is very close to experimental data in the same range of $m_{\pi\pi}$.
In order to improve the fit to the "up-flat" data we have used constraints given by the good fit to the "down-flat" data.
Two parameters were fixed by choosing the values of the scattering length and the slope parameter and two others by the values of phase 
shifts calculated from this fit at 500 and 550~MeV. 
A new fit with these constrains gave smaller value of $\chi^2_{exp}=13$ for 18 "up-flat" data points,
but simultaneously provided us with an enormous value of $\chi^2_{Roy} = 1.2\times 10^4$.
The phase shifts for this amplitude are presented in Fig.~2b by the dotted line.
It is clear that a simultaneous good fit to the "up-flat" data and to Roy's equations is impossible.

\begin{figure}[h!]
\includegraphics*[width=17.7pc]{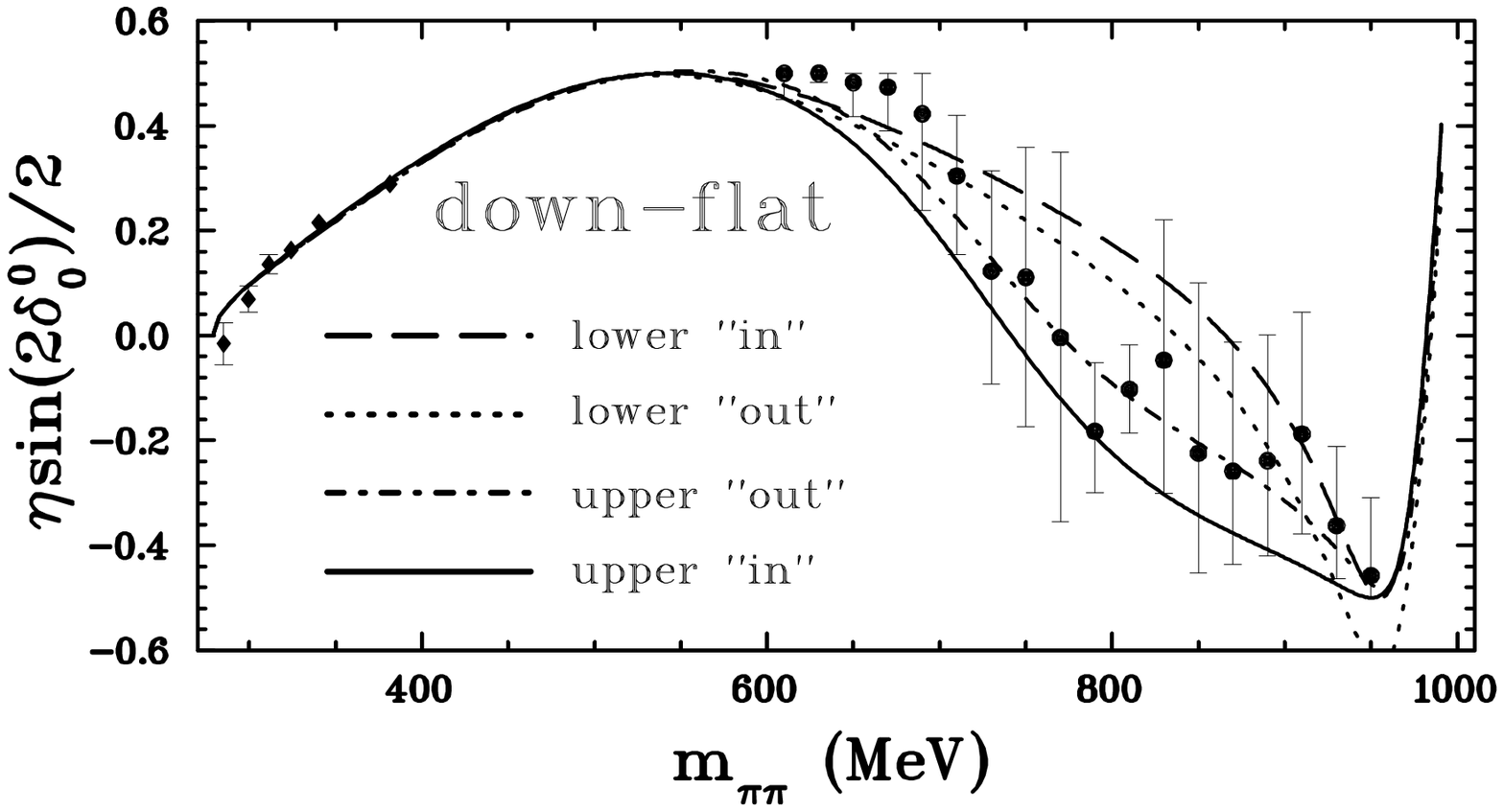}
\includegraphics*[width=17.7pc]{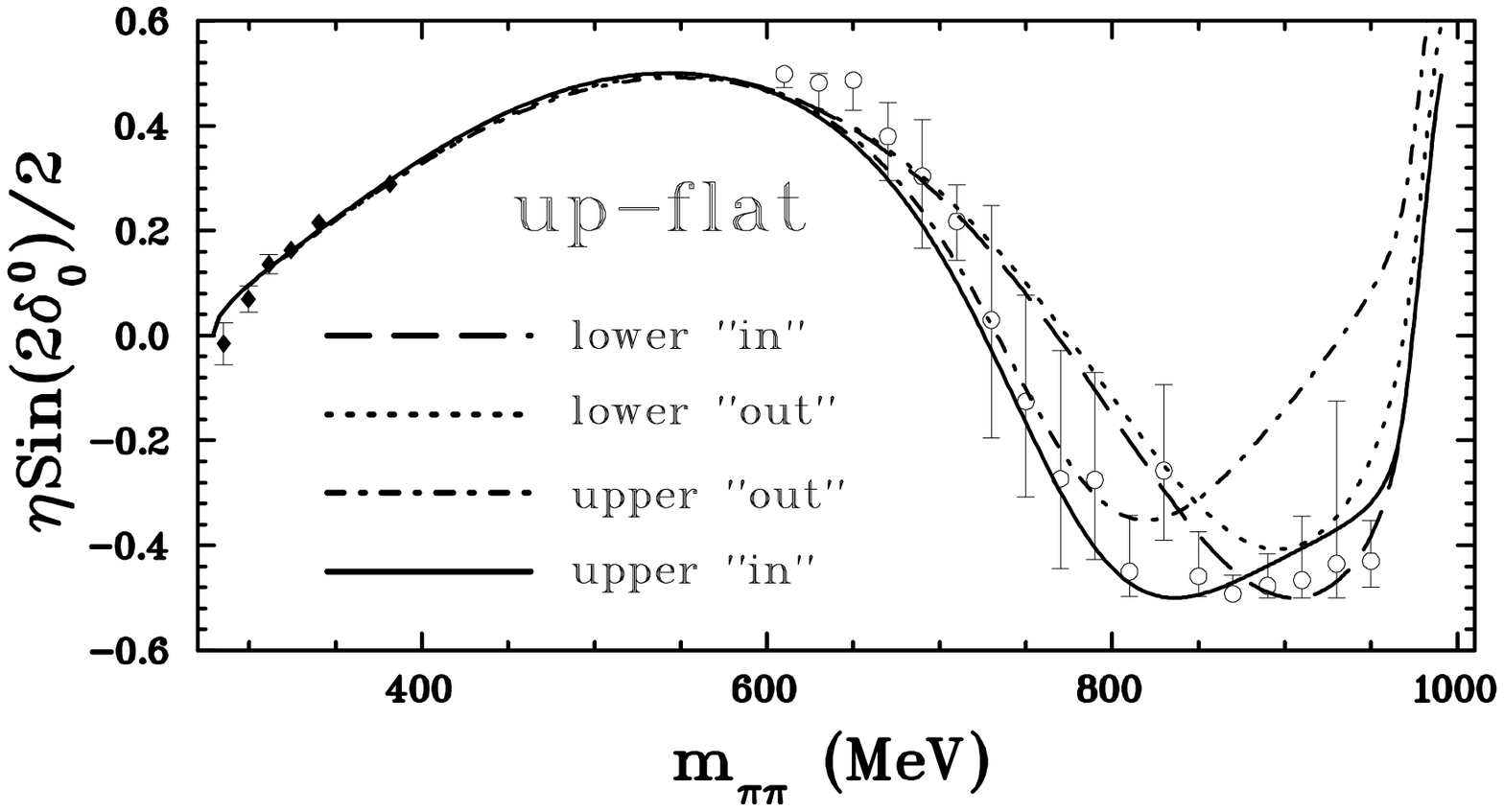}

\vspace{-0.5cm}

\caption{Real parts of input amplitudes ``{\it in}'' (their phase shifts have been presented in Fig.~2) and 
of output ones ``{\it out}'' calculated from Roy`s equations; all values are multiplied by $2\,k\,s^{-1/2}$.}
\label{fig:roydown}
\end{figure}
Besides the fits to the "up-flat" and "down-flat" experimental points we~have~also~performed fits to data points shifted upwards
and downwards by their errors.
In these fits the same four constraints, just described above, were used below 600~MeV.
Up to 937~MeV in the "down-flat" case in Fig.~3, the curves labeled {\it upper "in"} and {\it lower "in"} form a band included inside 
a band delimited by the lines {\it upper "out"} and {\it lower "out"}.
All the curves lying inside these bands correspond to the amplitudes fulfilling crossing symmetry so the "down-flat" data can
be accepted as physical ones.
In the "up-flat" case in Fig.~3 the output band lies outside of the input band from 840 to 970~MeV.
It means that in this case crossing symmetry is violated by the amplitudes fitted to the "up-flat" data.

\section{Fit to ``down-flat'' solution with uniform parameterizations}

Having chosen the ``down-flat'' solution as the physical one we have performed another fit  with different but 
uniform parameterizations in S0-, S2- and P1-waves separately (we have called this fit {\it uniform fit}). 
We have used our three coupled channel model \cite{kll97} for the S0- and S2-waves and constructed an analogous one
for the P1-wave.
The {\it uniform fit} has been done in wider  $m_{\pi\pi}$ range, than that described in section 2 (hereafter called {\it nonuniform fit}).
In the S0-wave case the fit was done from the $\pi\pi$ threshold to 1600 MeV and to 1250 MeV for the S2- and P1-wave cases.
In table 1 we present values of $\chi^2$ functions defined in equations (\ref{eq:chi:exp}) and (\ref{eq:chi:Roy}) for both fits.
The values of $\chi^2$ are bigger in the {\it uniform fit} due to a larger $m_{\pi\pi}$ range where the fit has been performed and to
a smaller flexibility of the model.


\begin{table}[hbt]
\caption{Values of $\chi^2$ functions for {\it nonuniform fit} (first line) and {\it uniform fit} (second line); 
in parentheses is number of experimental points; for $\chi^2_{tot}$ the second number is the total number of free parameters used in the fit.}
\label{tab:chi2val}
\begin{tabular}{ccc}
\hline
$\chi^2_{tot}$ & $\chi^2_{exp}$ & $\chi^2_{Roy}$   \\
43.9 (80/13) & 37.4 (44) & 6.5 (36) \\
247.5 (245/22) & 221.7 (179) & 25.8 (66)  \\
\hline
\end{tabular}
\end{table}


In table 2 we compare the values of the near-threshold parameters for both {\it nonuniform} and {\it uniform fits}.
Significant differences in the P1-wave case are due to the fact that near-threshold parameters of this wave are not sufficiently constrained 
both by Roy`s equations and by the experimental data.


\begin{table}[hbt]
\caption{Values of threshold parameters for S0, S2 and P1 waves for fits to ``down-flat'' solution. 
The first raw gives the results for the {\it nonuniform fit} and the second one those for the {\it uniform fit}.
In the S0- and S2-wave cases the $a_0^I$ and $b_0^I$ are multiplied by $m_{\pi}$ and  $10^2\,m_{\pi}^2$  respectively 
and in the case of the P1 wave by $10^2\,m_{\pi}^2$ and  $10^3\,m_{\pi}^4$. 
Parameters for the {\it nonuniform fit} with error bars are~presented~in~\cite{kll03}.}
\label{tab:thrpar}
\begin{tabular}{cccccc}
\hline
$a_0^0$ & $b_0^0$ & $a_0^2$ & $b_0^2$& $a_1^1$ & $b_1^1$ \\
0.22 & 0.25 & -3.4 & -7. & 4. & 2.6 \\
0.20 & 0.24 & -4.4 & -8. & 5. & 7.2 \\
\hline
\end{tabular}
\end{table}


\section{Conclusions}

The use of Roy`s equations allows final elimination of the long-standing up-down ambiguity in the $\pi\pi$ phase shifts in favour 
of the ``down-flat'' solution.
This agrees with the resent results of the joint $\pi^+\pi^-$ and $\pi^0\pi^0$ analysis \cite{kam02}.
Additional constraints from for example  Froissart Gribov sum rules or from ChPT are needed to fix the near threshold behaviour of the 
$\pi\pi$ amplitudes.

\end{document}